# CONDUCTANCE-BASED DYNAMIC CAUSAL MODELING: A MATHEMATICAL REVIEW OF ITS APPLICATION TO CROSS-POWER SPECTRAL DENSITIES

Inês Pereira[*][a], Stefan Frässle[a], Jakob Heinzle[a], Dario Schöbi[a], Cao Tri Do[a], Moritz Gruber[a], Klaas E. Stephan[a,b]

[a] Translational Neuromodeling Unit (TNU), Institute for Biomedical Engineering, University of Zurich & ETH Zurich, 8032 Zurich, Switzerland.

[b] Max Planck Institute for Metabolism Research, Cologne, Germany

**ABSTRACT**

Dynamic Causal Modeling (DCM) is a Bayesian framework for inferring on hidden (latent) neuronal states, based on measurements of brain activity. Since its introduction in 2003 for functional magnetic resonance imaging data, DCM has been extended to electrophysiological data, and several variants have been developed. Their biophysically motivated formulations make these models promising candidates for providing a mechanistic understanding of human brain dynamics, both in health and disease. However, due to their complexity and reliance on concepts from several fields, fully understanding the mathematical and conceptual basis behind certain variants of DCM can be challenging. At the same time, a solid theoretical knowledge of the models is crucial to avoid pitfalls in the application of these models and interpretation of their results. In this paper, we focus on one of the most advanced formulations of DCM, *i.e.* conductance-based DCM for cross-spectral densities, whose components are described across multiple technical papers. The aim of the present article is to provide an accessible exposition of the mathematical background, together with an illustration of the model's behavior. To this end, we include step-by-step derivations of the model equations, point to important aspects in the software implementation of those models, and use simulations to provide an intuitive understanding of the type of responses that can be generated and the role that specific parameters play in the model. Furthermore, all code utilized for our simulations is made publicly available alongside the manuscript to allow readers an easy hands-on experience with conductance-based DCM.

[*] Corresponding author at: Translational Neuromodeling Unit (TNU), Institute for Biomedical Engineering, University of Zurich & ETH Zurich, Wilfriedstrasse 6, 8032 Zurich, Switzerland.

*Email address*: pereira@biomed.ee.ethz.ch

# 1 Introduction

Dynamic Causal Modeling (DCM) is a framework to construct generative models describing how putative neural mechanisms give rise to neurophysiological data. By inverting these models, using generic Bayesian techniques, it is possible to infer upon hidden (unobserved) neuronal states from measured data; for reviews, see [1]–[4]. As a modeling technique, DCM has been used to study physiological processes in the healthy human brain (for examples, see [5]–[7]). In addition, it has also been increasingly employed in the fields of Computational Psychiatry [8]–[13] and Computational Neurology [14]–[19], with the hope of not only providing mechanistic insights into pathophysiology, but also of developing the model into a useful clinical tool (*i.e.*, a "computational assay" [20], [21]).

DCMs[1] are formulated using ordinary or stochastic differential equations which describe the dynamics of neural (hidden) states over time [22]. First proposed in 2003 by Friston *et al.* [23], DCM was initially developed for functional magnetic resonance imaging (fMRI) data. In DCM for fMRI, the neuronal model is simple, assuming a single state variable per region and modelling neuronal population dynamics by using a bilinear or $2^{nd}$-order Taylor approximation [24]. This relatively simple model is used to estimate how specific brain regions interact with each other through directed synaptic interactions (effective connectivity) and to evaluate whether and how this connectivity is influenced by experimentally controlled factors and/or disease processes. In 2006, David *et al.* [25] extended this modeling technique to event-related responses (ERPs) measured with electroencephalography (EEG) or magnetoencephalography (MEG). In this seminal work, DCM of ERPs was cast in terms of a neural mass model based on the Jansen-Rit model [26] which represents a more sophisticated model of neuronal dynamics as compared to the relatively abstract descriptions in DCM for fMRI. This biological realism is afforded by the richer temporal information contained in electrophysiological measurements (on the order of milliseconds), as compared to the coarse nature of the BOLD response (on the order of seconds) [27].

In 2009, Marreiros *et al.* [28] introduced an important variant of DCM for electrophysiological data, describing a conductance-based model based on the Morris–Lecar model [29]. This conductance-based DCM (cbDCM) contains explicit representations of ionotropic receptors with distinct time constants, namely the AMPA, NMDA and $GABA_A$ receptors. Given the central role of these ionotropic receptors in many psychiatric and neurological disease processes and given that drugs targeting NMDA and $GABA_A$ receptors exist, this mechanistically fine-grained formulation is of considerable interest for studying neural circuits and their alterations in disease and under pharmacological interventions. For example, cbDCM has been applied to data from patients with monogenic channelopathies [15] and NMDA receptor antibody encephalitis [14] as well as to data from pharmacological studies in healthy volunteers and animals [30]–[32].

Unfortunately, the literature introducing cbDCM and its mathematical foundations is distributed over several papers and is not easily accessible for the average neuroscientist or clinician. This tutorial-style paper offers a didactic treatment of the model, focusing on the underlying theory and mathematical derivations. These derivations are here expanded beyond what is presented in the original papers, in order to provide a more detailed, step-by-step description of the model. In addition, several footnotes will be added along the way, so as to offer extra hints without breaking the flow of the manuscript. Furthermore, we will discuss practical aspects related to the inversion of DCMs using Statistical Parametric Mapping (SPM), a freely available open-source and widely used software package written in MATLAB [33]. Notably, this paper is not meant to provide an exhaustive review of the existing literature on variants of DCM for EEG/MEG. Comprehensive reviews on this topic as well as general practice recommendations can be found in other publications [1], [2], [34]. Finally, this paper will not cover the topics of Bayesian inference or model inversion in detail. More in-depth accounts of these concepts can be found elsewhere [35], [36].

---

[1] In this paper, we used the acronym "DCM" both to refer to the modeling approach (dynamic causal modeling) and to its instantiation (dynamic causal model).

Generally, a DCM for electrophysiological data comprises two parts (Figure 1): the **neuronal model**, which delineates the intra- and inter-neuronal source[2] dynamics, and the **observation model**, which describes how source activity propagates through surrounding tissues (brain, skull, scalp) in order to generate the data registered at the level of the sensors [37]. In this paper, for conceptual clarity, we will first review how to characterize and distinguish the existing variants of DCM for EEG/MEG and local field potentials (LFP). In doing so, we will provide insights as to which model could be used, depending on the research question considered. Secondly, we offer an introduction to conductance-based models. We outline the neuronal model of cbDCM, moving from the single-neuron level to the population level, to the level of the cortical column, before finally describing connectivity amongst sources. In addition, we give an account of the observation model for cross-spectral densities. In parallel, we review the literature where this model was introduced and described. However, since these seminal papers were published, the implementation of these models in SPM has undergone several refinements and modifications. Therefore, we also review some of the newer aspects of SPM that are relevant for the user.

The paper assumes that the reader has basic knowledge of neuroanatomy, neurophysiology, Bayesian statistics, and signal processing techniques such as the Fourier transform and convolutions. A list of important concepts, along with their definitions, is provided in Table 1.

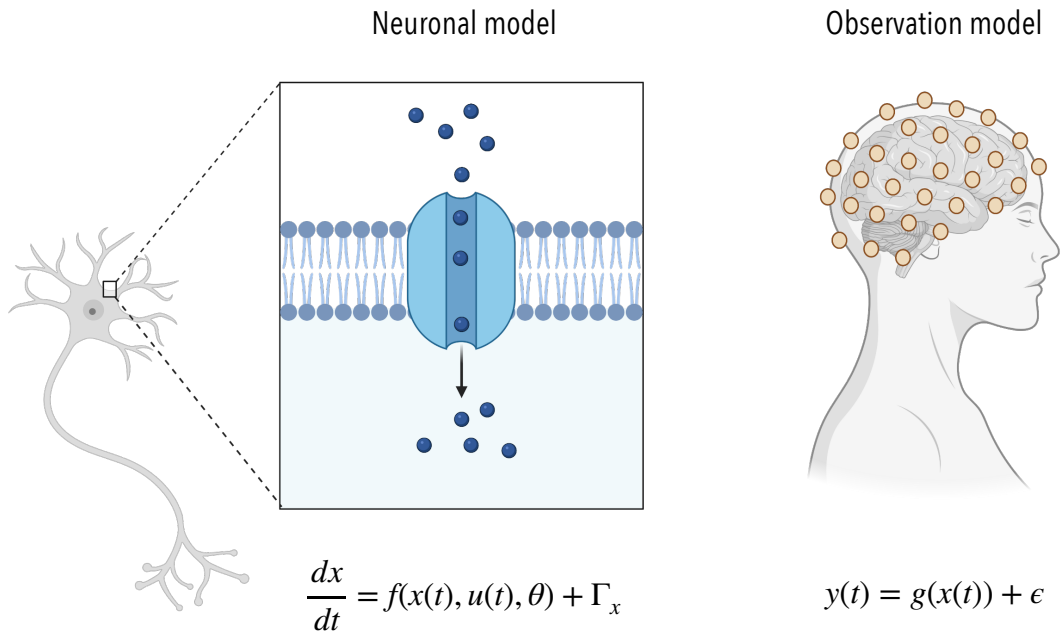

Figure 1: General structure of DCMs. A DCM for electrophysiological data comprises two parts: (1) the **neuronal model**, which delineates the dynamics $f(x(t), u(t), \theta)$ of the hidden states $x(t)$, as well as ensuing neuronal population activity, and (2) the **observation model**, which describes how source activity $g(x(t))$ propagates through surrounding tissues (brain, skull, scalp) in order to generate the data $y(t)$ registered at the level of the sensors. $u(t)$ are the inputs to a neuronal population, $\theta$ represents the model parameters and $\Gamma_x$ and $\epsilon$ are stochastic noise terms. Figure created with Biorender.com.

---

[2] In this context, a "source" refers to a neuronal population (or region) of the brain. This terminology is common because activity in a neuronal population constitutes the *source* of signal.

*Table 1: Important concepts and definitions*

| Concept | Definition |
|---|---|
| *Statistic* | Index of some attribute of given data (*e.g.,* sample mean). "Moment" is sometimes used interchangeably with "statistic", where the first moment corresponds to the mean and the second moment to the covariance. |
| *Time series* | *"A series of values of a quantity obtained at successive times, often with equal intervals between them"* [38]. |
| *Resting-state* | A condition of unconstrained cognition in which there is no exogenous, experimental input and brain activity is therefore "spontaneous". |
| *Sensor* | *"Device which detects or measures a physical property and records, indicates, or otherwise responds to it"* [39]. In the context of this paper, it is used interchangeably with "channel" or "electrode". |
| *Frequency band* | *"Range of frequencies […] between two limits"* [40]. The main frequency bands used in EEG analysis include the delta (<4 Hz), theta (4-7 Hz), alpha (8-12 Hz), beta (13-29 Hz) and gamma (30-79) bands [41]. |
| *Hidden state* | Unobserved quantity. "Hidden state" is used interchangeably with "latent state". In deterministic forms of DCMs (where trajectories of states are fully determined by the values of the model's parameters and [known] inputs), we wish to model the neuronal hidden states by inferring upon the model parameters using observable data. |
| *Density* | In the context of this publication, the term "density" (*e.g.*, as in "ensemble density") refers to the probability density function of a continuous random variable. |
| *Generative model* | A probabilistic model that describes the putative process by which data were generated. Mathematically speaking, generative models specify the joint probability density (product of likelihood and prior) over model parameters and measured data. By sampling from the prior, it is possible to generate synthetic data points [36]. |
| *Forward model* | Mapping from hidden states (*e.g.*, neuronal activity) to observed data (*e.g.*, EEG measurements). "Forward model" will be used interchangeably with "observation model" [34]. A neuronal model and an observation model together make up a DCM. |
| *Model inversion* | In a Bayesian setting, the process by which a generative model is used to compute the posterior distribution of the model parameters. |
| *Mean field model (MFM)* | A population (ensemble) of spiking neurons can be modeled using a population density function which describes the probabilistic evolution of the population response over time. Such a mean field model (MFM) is described by the dynamics of the moments of the population density function [34].<br><br>MFMs have been used for more than half a century and are defined using concepts from statistical physics (refer to the Fokker-Planck equation below) [42], [43]. Since EEG/MEG data reflect the activity of populations of neurons, MFMs are well suited for these data modalities. |
| *Neural mass model (NMM)* | If one only considers the first moment (*i.e.*, the mean) of a MFM, one obtains a neural mass model (NMM) [43]. NMMs are therefore a special case of MFMs [34]. |
| *Neural field model (NFM)* | NMMs and MFMs consider the evolution of the neuronal states only over time. However, this evolution can also be modeled over space (*e.g.,* across the cortical sheet). Models involving differential operators with both temporal and spatial terms are called neural field models [43]. |
| *Fokker-Planck equation* | The Fokker-Planck equation is a partial differential equation from statistical physics that allows one to describe, using a flow-diffusion process, the evolution of a probability density function of an ensemble of individual components (*e.g.*, neurons) over time. The advantage offered by this approach is that it permits modeling of ensemble density dynamics in a deterministic manner, even if the dynamics of individual components are stochastic. |

## 2  DCM variants

DCMs for electrophysiological data come in two different flavors: neurophysiological and phenomenological. Phenomenological models include DCM for induced responses and DCM for phase coupling [44], [45], and are characterized by an evolution function of the neuronal states that is not closely related to the underlying neurophysiology. These models will not be addressed in this paper; however, for details, please consult the relevant literature [44], [46]. The focus of the present paper will therefore be on neurophysiologically-informed DCM variants, which can again be divided into several categories depending on several dimensions (compare Figure 2):

- **How the cortical column is described** — We distinguish "convolution-based" models, which primarily consider the cortical column, from "conductance-based" variants, which start by modeling a single cell's electrophysiological properties [34]. In addition, there are different ways of modeling a cortical column. The first DCMs for electrophysiological data were based on the Jansen-Rit model [26] and included 3 neuronal populations (excitatory pyramidal cells, excitatory spiny stellate cells, and inhibitory interneurons), whilst the more recent "canonical-microcircuit" variant takes 4 populations into account (superficial and deep pyramidal cells, spiny stellate cells and inhibitory interneurons) [34].

- **How the hidden states of the neuronal populations are modeled** — *i.e.*, whether a population's density is summarized by a single number (first-order statistic), as is the case for neural-mass models (NMM); or whether higher-order statistics are also taken into account, as is the case for mean-field models (MFM). If one makes the neuronal states not only a function of time (as in the NMM and MFM), but also a function of space, one obtains a neural-field model (NFM).

- **Whether or not there is an exogenous/experimental input** — Three types of neuronal activity can be captured: **1)** Event-related potentials (ERP), which correspond to the phase-locked response of a neural system to a stimulus and can be characterised by averaging in the time domain [47]. **2)** Induced responses, which constitute changes in neuronal oscillations that appear after a stimulus, but are not phase-locked to this stimulus; they can be obtained by averaging in the frequency domain [47], [48]. As explained before, DCM for induced responses [46] will not be discussed in detail in this paper. **3)** "Resting-state" or "spontaneous" activity, where no exogenous (experimental) input is present. In the latter case, the output is summarized in the frequency domain, *i.e.*, we obtain resting-state spectral oscillations.

- **The type of data modeled** — *i.e.*, EEG data, MEG data, or local field potentials (LFP).

Importantly, the defining aspects considered above can be combined. For example, it is possible to apply neural-mass or mean-field formulations to convolution or conductance-based models [34]. Choosing an adequate DCM variant involves careful consideration of the data and problem at hand. This is because different (clinical) questions might be more naturally addressed by some variants than others. For instance, in an autoimmune disorder called anti-NMDA receptor encephalitis, auto-immune antibodies selectively target the NMDA receptor, leading to receptor hypofunction and a variety of severe psychiatric and neurological conditions, including psychosis and epilepsy. Hence, in order to test disease-related hypotheses about NMDA receptor function directly, it may be advisable to use a DCM that can represent this receptor explicitly, such as cbDCM. This was precisely the modeling approach followed by Symmonds *et al.* [14] who used cbDCM to model EEG data from anti-NMDA receptor encephalitis patients and controls. Hence, the choice of the optimal DCM variant has to be tailored to the specific hypothesis about (disease-relevant) processes of interest.

In the following, we define what constitutes a conductance-based model.

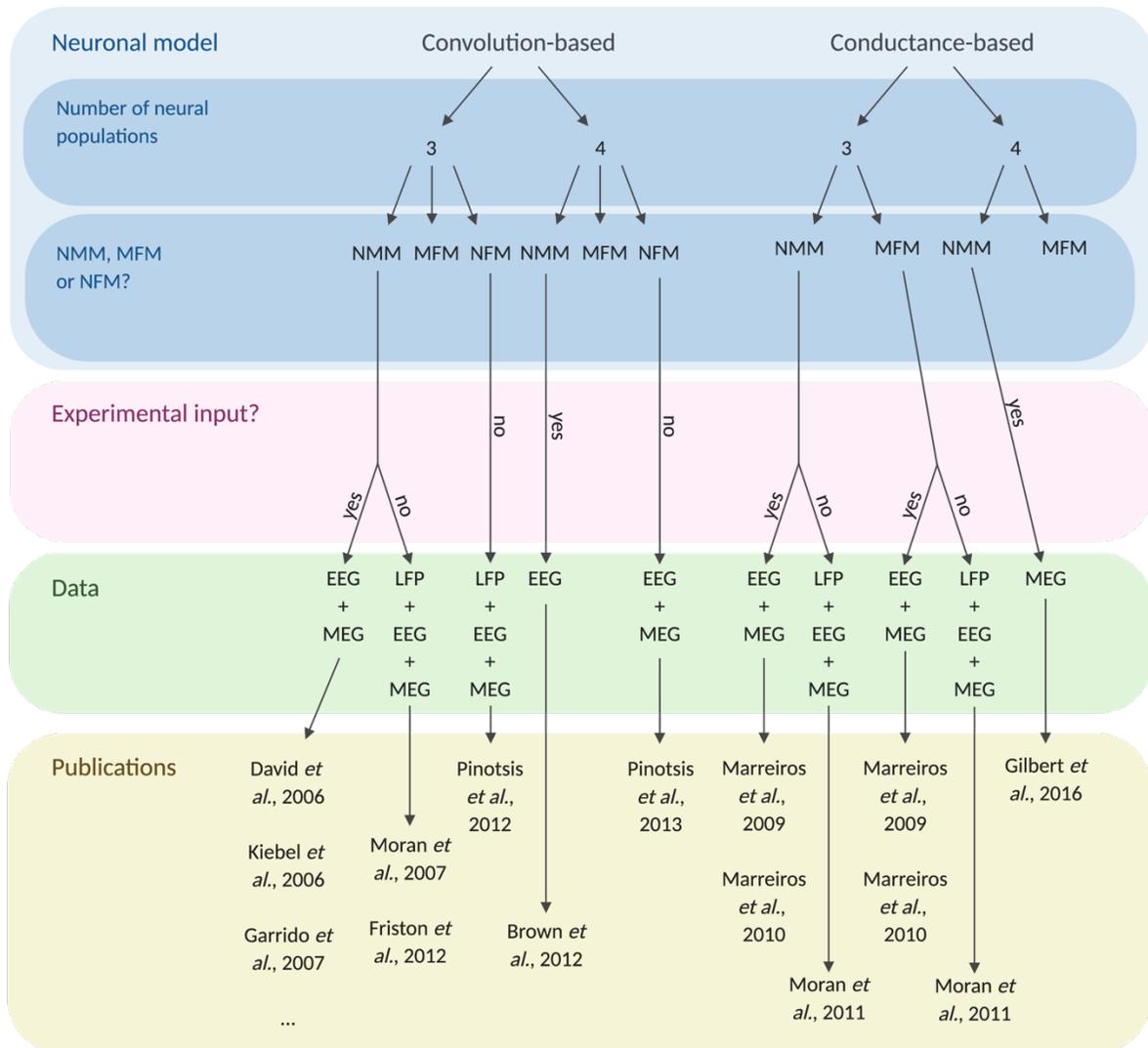

*Figure 2: DCM variants, along with examples of relevant publications* [15], [25], [28], [49]–[57]. *Models of induced responses are not considered. EEG: electroencephalography; LFP: local field potential; MEG: magnetoencephalography; MFM: mean field model; NFM: neural field model; NMM: neural mass model. Figure created with Biorender.com.*

## 3   Conductance-based models

The electrical properties of neurons can be described by the membrane conductance associated with different ions. In 1952, Hodgkin and Huxley [58] famously used this formalism to describe how sodium and potassium currents can generate action potentials in the giant axon of the squid [59]. It has been extended to include other ions (see Morris and Lecar, 1981 [29]) and can accommodate active, neurotransmitter-mediated ion flow, as well as leaky ion channels and externally applied current [60]. Conductance-based models are hence models of excitable cells (*e.g.*, neurons) that represent ion channels through their conductance [61]. According to Dayan and Abbott (2001), these models have been shown to "*reproduce the rich and complex dynamics of real neurons quite accurately*" [59].

In the following sections, we will describe the conductance-based neuronal model implemented in DCM for electrophysiological data. In our description, we will move from the single-neuron level, to the level of the cortical column, and finally describe macroscopic source connectivity.

## 4 Modeling a single neuron

Conductance-based models describe how ions flow into and out of excitable cells via a parallel resistor-capacitor (RC) circuit. Using these so-called equivalent circuit models, it is possible to model the dynamics associated with specific ions or specific ligand-gated ion channels (also referred to as ionotropic receptors). Let us consider a neuron with two glutamate receptors, AMPA and NMDA, and the GABA$_A$ receptor[3] (see Figure 3). Note that both the AMPA and NMDA receptors are permeable to multiple cations, not just sodium. A leak current $L$ is also included, to model the effect of other passive ion channels on the cell's resting membrane potential.

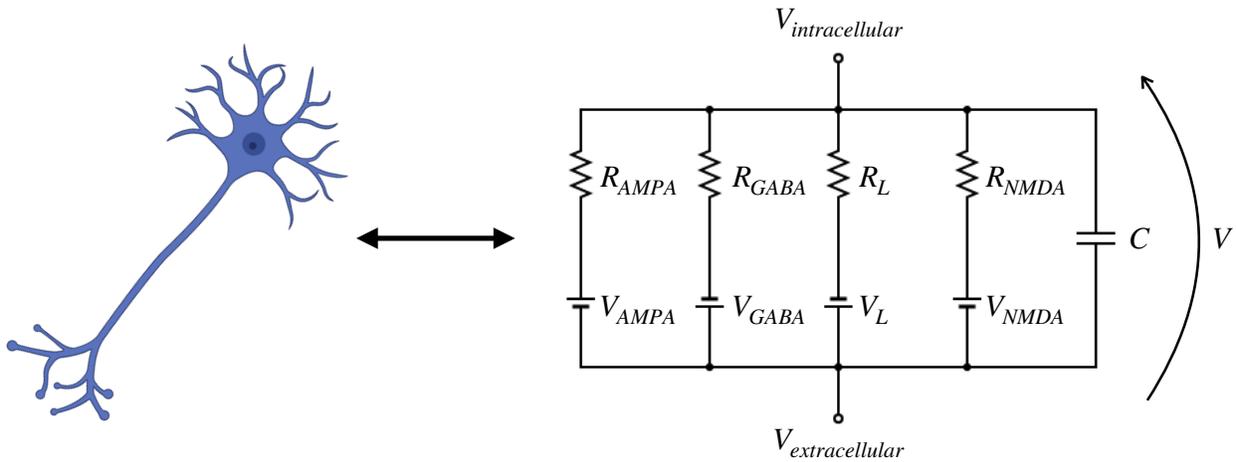

*Figure 3: Equivalent circuit model for a neuron. $V_{intracellular}$ and $V_{extracellular}$ indicate the intra- and extracellular potentials. The difference between these two quantities is called the membrane potential V, transmembrane potential or membrane voltage. $R_k$ indicates the resistance associated with ion channel $k$: either AMPA, GABA, NMDA or L. L denotes the passive leak current channels. Finally, $V_k$ represents the reversal potential for ion channel $k$. Figure created with Biorender.com*

The equation of motion for the membrane potential *V* can be derived by making use of two fundamental laws of physics. Specifically, we make use of Kirchhoff's current law [60]:

$$u = I_c + \sum_k I_k, \qquad (1)$$

---

[3] AMPA, NMDA and GABA$_A$ receptors are associated with distinct time constants, a condition necessary for identifying their relative contribution to the measured potentials.

where $u$ is the injected current, $I_c$ the capacitive current, and $I_k$ is the resistive current associated with the $k$-th channel. Re-arranging this equation yields:

$$I_c = -\sum_k I_k + u \tag{2}$$

We further use the current-voltage relation of a capacitor to re-express the capacitive current $I_c$:

$$I_c = C \cdot \frac{dV}{dt} \tag{3}$$

In addition, we apply Ohm's law to express the resistive current associated with each channel $k$:

$$I_k = \frac{V - V_k}{R_k} = g_k(V - V_k) \tag{4}$$

Here, $V_k$ represents the reversal potential for channel $k$ and the conductance $g_k = 1/R_k$ is given by the inverse resistance $R_k$. By combining Equations 2, 3 and 4, we can relate the membrane capacitive current to the cellular ionic currents.

$$C\dot{V} = \left(\sum_k g_k (V_k - V)\right) + u \qquad k \in L, \text{AMPA, GABA} \tag{5}$$

Note, however, that Equation 5 does not yet contain a term that explicitly represents the NMDA receptor. This requires a slight extension because of the so-called voltage-dependent magnesium ($Mg^{2+}$) block of the NMDA receptor. Indeed, at hyperpolarized potentials, magnesium is present within the channel pore, effectively blocking it (Figure 4, left). Depolarization of the membrane potential pushes $Mg^{2+}$ out of the pore, allowing current to flow through the channel [62].

This $Mg^{2+}$ nonlinearity is described as follows [50], [60], [63]:

$$m(V) = \frac{1}{1 + 0.2 \cdot exp(-\alpha_{NMDA} \cdot V)} \tag{6}$$

where $\alpha_{NMDA}$ represents the magnesium block parameter[4] (Figure 4, right).

---

[4] In the current implementation of SPM12: $\alpha_{NMDA}$ is fixed and set to 0.06 (see mg_switch.m).

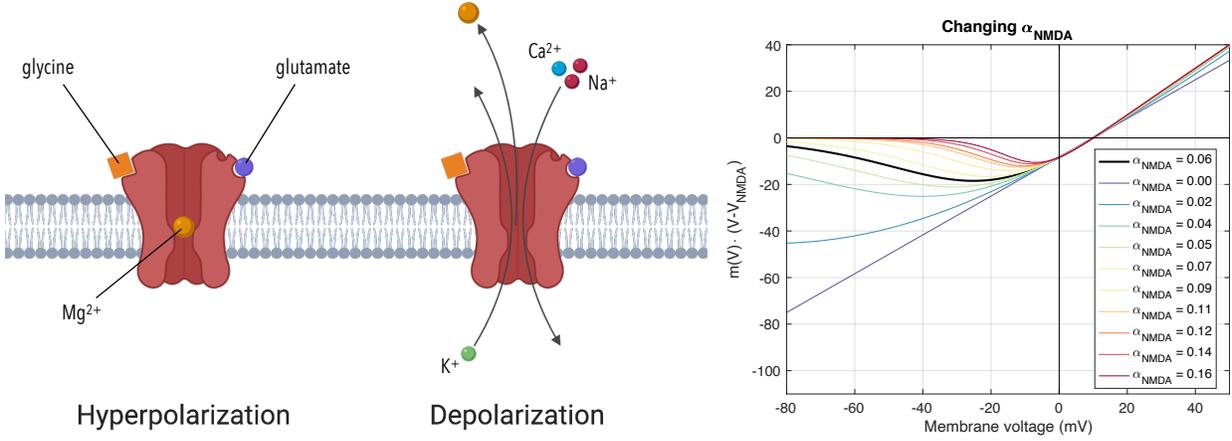

*Figure 4: **(Left)** Voltage-dependent block of the NMDA receptor pore by $Mg^{2+}$. Note how channel opening necessitates neurotransmitter binding as well as depolarization of the cell. Figure adapted from [62]. **(Right)** $m(V)(V - V_{NMDA})$ represented as a function of membrane voltage. The black line depicts the behavior of $m(V)(V - V_{NMDA})$ for the parameter value currently used in SPM12. Figure created with Biorender.com*

By adding a term that represents the NMDA receptor, as well as a noise term $\Gamma_V$, the final expression for the equation of motion for the membrane potential $V$ is obtained [50]:

$$C\dot{V} = \left(\sum_k g_k (V_k - V)\right) + g_{NMDA} m(V)(V_{NMDA} - V) + u + \Gamma_V \qquad k \in L, \text{AMPA, GABA} \qquad (7)$$

Where:

- $C$ is the membrane capacitance;

- $V$ the membrane potential;

- $\dot{V}$ the time derivative of $V$;

- $g_k$ represents the conductance for channel $k$;

- $V_k$ is the reversal potential for channel $k$;

- $u$ is the applied input current. For cells which receive no external input: $u = 0$;

- $\Gamma_V$ is a stochastic term which models Gaussian noise;

- $L$ denotes the passive leak current channels.

We hence have our first set of differential equations, which model the change in membrane **voltage** of a single neuron over time. Now, we turn to the channel conductances. The leak channel conductance $g_L$ is assumed to be fixed. Thus, we define the subsequent equation only for $k \in$ {AMPA, $GABA_A$, NMDA} [50]:

$$\dot{g}_k = \kappa_k(\gamma_{aff} \cdot \sigma_{aff} - g_k) + \Gamma_g \qquad (8)$$

where $\kappa_k$ is the inverse time constant for the $k$-th receptor, $\sigma_{aff}$ models the firing from afferent neurons weighted by $\gamma_{aff}$, and $\Gamma_g$ represents the stochastic component, again modeling Gaussian noise.

In short, the model entails **four hidden states**: the membrane voltage $V$ and the conductance $g_k$ for all three receptors: AMPA, GABA$_A$ and NMDA.

## 5   Modeling a population of neurons

Let us start by looking at the equations of motion for the hidden states $x$ from the previous section (Equations 7 and 8). Note that they are stochastic differential equations of the form (Figure 1):

$$\dot{x} = f(x) + \Gamma_x \qquad (9)$$

Here, $f(x)$ corresponds to the deterministic part of the equation and $\Gamma_x$ is the stochastic component. Neurobiologically, this stochastic component represents different sources of noise. An example at the single-neuron level is *thermal noise, i.e.,* thermal fluctuations which trigger spontaneous conformational changes in proteins that are part of the ion channels, some of which lead to opening or closing of the channel (*i.e., channel noise*)[5]. For an extensive review on the sources of noise in the brain, refer to [64], [65].

At this point, we are still at the level of single neurons. However, techniques like EEG or MEG cannot resolve the behavior of single neurons. Instead, they provide measurements of the activity of large **populations (or ensembles)** of neurons. We thus focus on the joint behavior of the population by characterizing the evolution of the ensemble density of all neurons in state space (*i.e.*, the space spanned by the ensemble's voltage and conductance). The dynamics of this ensemble density follow the Fokker-Planck equation [66] (Table 1), a concept from statistical physics. The strength of this approach is that it enables us to reformulate the stochastic single-neuron dynamics as deterministic equations for the whole population of neurons. This is done by describing the temporal evolution of the ensemble density as a flow-diffusion process (Table 1).

If we assume that this ensemble density can be represented by a Gaussian distribution $q(x) = \mathcal{N}(\mu, \Sigma)$ (the so-called Laplace approximation), these equations can be reformulated as ordinary differential equations of the sufficient statistics of the population's density. For each hidden state $l$ in the $j$-th neuronal population, the equations that describe the dynamics of the sufficient statistics are [28], [50]:

$$q(x) = \mathcal{N}(\mu, \Sigma) \qquad (10)$$

---

[5] Later, we will discuss how to model the so-called neuronal innovations, which represent spontaneous endogenous fluctuations of the neuronal signal at the level of a network of sources.

$$\dot{\mu}_l^{(j)} = f_l^{(j)}(\mu) + \frac{1}{2}Tr\left(\Sigma^{(j)}\frac{\partial^2 f_l^{(j)}}{\partial x^2}\right) \tag{11}$$

$$\dot{\Sigma}^{(j)} = \frac{\partial f^{(j)}}{\partial x}\Sigma^{(j)} + \Sigma^{(j)}\frac{\partial f^{(j)T}}{\partial x} + D^{(j)} + D^{(j)T} \tag{12}$$

With Equations 10-12, we have thus defined a mean-field model (MFM), where, $f_l^{(j)}$ is the deterministic part of the $j$-th neuronal population's $l$-th hidden state value, and $D$ represents the so-called diffusion coefficient. From Equation 11, we can see that in this mean-field formulation the covariance matrix $\Sigma^{(j)}$ affects the dynamics of the mean of the ensemble density. By contrast, in neural-mass models (NMMs), this interdependence between mean and variance is dropped, which means that the product of the Hessian $\partial^2 f_l^{(j)}/\partial x^2$ and the covariance matrix $\Sigma^{(j)}$ or the trace of the product has to evaluate to zero (for details, consult [28]). While NMMs enjoy much attention because of their relative simplicity, MFMs are capable of representing more complex dynamics, including oscillatory signals [28]. In the following equations, we consider the NMM formulation for cbDCM.

In this case, one simply recovers the deterministic part of Equations 7 and 8 [28]:

$$C\dot{\mu}_V^{(j)} = \left(\sum_k \mu_{g_k}^{(j)}\left(V_k - \mu_V^{(j)}\right)\right) + \mu_{g_{NMDA}}^{(j)}m\left(\mu_V^{(j)}\right)\left(V_{NMDA} - \mu_V^{(j)}\right) \quad k \in L, \text{AMPA, GABA} \tag{13}$$

$$\dot{\mu}_{g_k}^{(j)} = \kappa_k^{(j)}\left(\zeta_k^{(j)} - \mu_{g_k}^{(j)}\right) \tag{14}$$

$\mu_V^{(j)}$ is hence the $j$-th population's mean voltage, whereas $\mu_{g_k}^{(j)}$ is the population mean conductance for channel $k$. We can now express $\zeta_k^{(j)}$, the input to the $j$-th neuronal population, as:

$$\zeta_k^{(j)} = \sum_i \gamma_k^{(j,i)}\sigma\left(\mu_V^{(i)} - V_R, \Sigma^{(i)}\right) \tag{15}$$

where $\gamma_k^{(j,i)}$ represents a coupling parameter for channel type $k$, from population $i$ to population $j$. Moreover, $\sigma(\cdot)$ is the cumulative distribution function of the univariate normal distribution $\mathcal{N}\left(\mu_V^{(j)} - V_R, \Sigma^{(j)}\right)$. Intuitively, $\sigma(\cdot)$ can be understood as the proportion of active (spiking) afferent neurons. Here, the variance $\Sigma^{(j)}$ is also a free parameter [50].

## 6  Modeling a source

Now that we are able to model the activity of single neuronal **populations** (as a reminder, we are not modeling single neurons but **ensembles** of neurons), we can define a cortical column (*i.e.*, a cortical functional unit).

The original formulation of DCM for EEG by David *et al.* [25] was based on the model by Jansen and Rit [26], which was constructed based on previous experimental work with cats and humans [67]. This initial formulation considered three different neuronal populations: excitatory pyramidal cells, excitatory spiny stellate cells and inhibitory interneurons (Figure 5). Spiny stellate cells are found in layer IV of the cortical column, whereas the other two cell populations are considered to occupy both the supra- and infragranular layers. Subsequently, a canonical-microcircuit model for DCM was developed [68], which further divides the pyramidal-cell population into two distinct superficial (supra-granular) and deep (infra-granular) subgroups.

The cell populations within a source interact through a set of intrinsic (or within-source) connections, represented in Figure 5. These connections are defined based on the Jansen-Rit model [26], in case of the 3-population model. The intrinsic connections are encoded by the coupling parameters $\gamma_k^{(j,i)}$ from Equation 15. By fully defining a layered circuit, with a specific set of neuronal populations and intrinsic connections, we have effectively described **one source.**

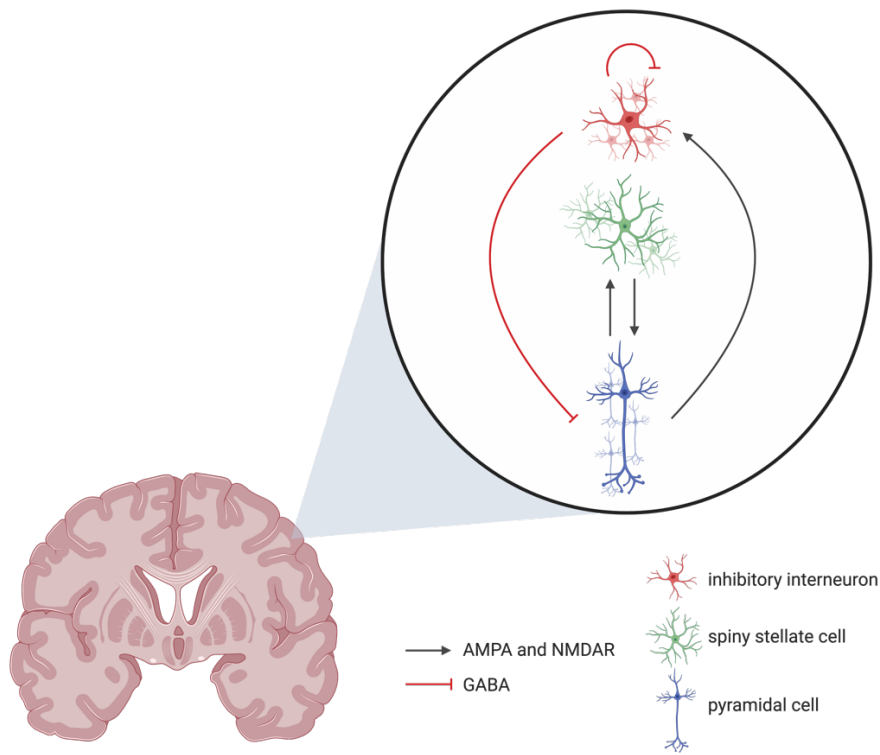

*Figure 5: Cortical column, based on the Jansen-Rit model* [26]. *Spiny stellate cells are found in layer IV of the cortical column, whereas the other two cell populations are considered to occupy both the supra- and infragranular layers. Note that we model populations of cells, not individual neurons. For illustration purposes, the intrinsic glutamatergic connections mediated by the AMPA and NMDA receptors are grouped. However, these can be modeled separately. Figure modified from Moran et al., 2013* [34]. *For the current implementation in SPM12, refer to Section 11. Figure created with Biorender.com*

## 7 Modeling between-source connectivity

In the previous section, we have explained how one can model a single source. This source is then set to represent a specific brain region: for instance, the medial prefrontal cortex or a particular visual area. However, neural processes typically unfold as the result of interactions in a network of multiple sources. As can be seen in Figure 6, such a network perspective can be taken by placing sources in several brain areas, which are then connected via weighted extrinsic or between-source connections.

In DCM for electrophysiological data, the definition of the between-source connections follows a simplified version of the connectivity rules proposed by Felleman and Van Essen in 1991 [69]. These were derived from experimental studies, most prominently on the monkey visual cortex. Extrinsic connections are divided into several types, based on the layers in which they terminate. **Forward connections**, which run from hierarchically lower to higher areas and mainly originate from supragranular layers, terminate in (granular) layer IV, whilst **backward connections**, which run from hierarchically higher to lower brain areas and mostly originate from infragranular layers, terminate in the supra- and infragranular layers, avoiding layer IV. **Lateral connections** terminate in all three layers (see Figure 7) [25], [70].

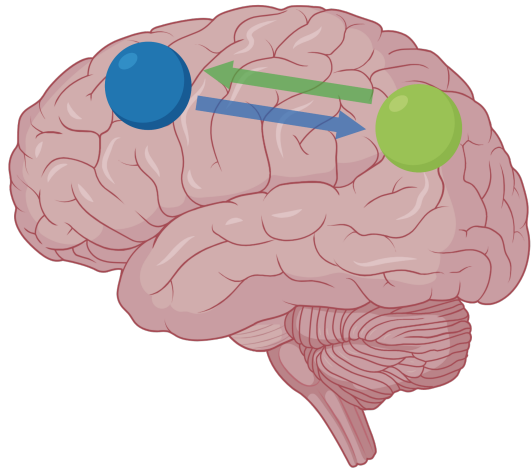

*Figure 6: Example locations for modeled sources, connected via extrinsic, or between-sources connections. Figure created with Biorender.com*

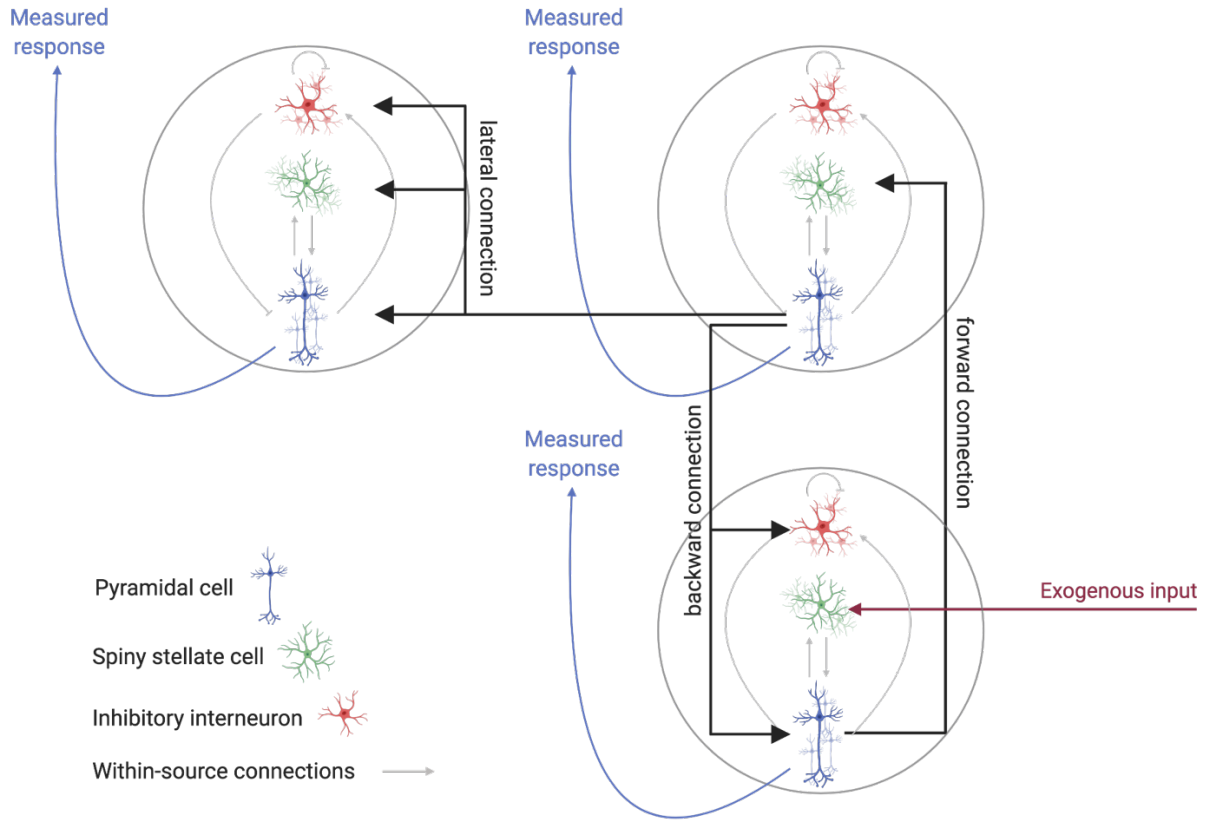

*Figure 7: Depiction of the several types of extrinsic, or between-source connections. Please note that the vertical arrangement of the 3 types of neurons should not be misinterpreted as statements about layers. While spiny stellate cells are only present in the granular layer (layer IV), pyramidal cells and inhibitory interneurons are located in both supragranular layers (layers II/III) and infragranular layers (layers V/VI). It is assumed that the exogenous input enters a source via the spiny stellate cells in the granular layer (layer IV). In addition, given that the apical dendrites of cortical pyramidal cells are mostly arranged perpendicularly to the cortical surface and in parallel with respect to each other, it is considered that pyramidal cells are the main generators of the measured M/EEG response [71]. For LFP, it is the arrangement and orientation of the dendritic tree versus the soma, along with the parallel arrangement of the apical dendrites, that make pyramidal cells the main generators of signal. As such, in DCM for electrophysiological data, pyramidal cells are assumed to have the most important contribution to the measured signal. Figure created with Biorender.com*

## 8  Observation model

To model how the dendritic signal from the pyramidal cells gives rise to the measured signal in our sensors (*e.g.,* EEG electrodes), another layer needs to be defined on top of our neuronal model. This observation model $g(x(t), \theta)$ is, in principle, nothing more than a projection from sources to the sensors. This projection is assumed to be linear and instantaneous and can be expressed as follows [72]:

$$\hat{y}_i(t) = g(x(t), \theta) = L(\theta)x(t) \tag{16}$$

Where $\hat{y}_i(t)$ represents the **predicted measurements** at sensor $i$, $g(x(t),\theta)$ the **observation model** with parameters $\theta$, $x(t)$ is the **pyramidal cell activity** and $L(\theta)$ the **lead field or gain matrix**. This matrix describes the passive conduction of the electromagnetic field from sources to sensors and assumes different forms depending on the modality considered (*i.e.*, EEG, MEG or LFP). To complete the model of time series data $y_i(t)$, one now only needs to furnish the model with assumptions about the measurement noise $\epsilon \sim \mathcal{N}(0,\Sigma)$:

$$y_i = \hat{y}_i + \epsilon \tag{17}$$

In ERP models of LFP or M/EEG data, predictions and data are dealt with in the time domain. However, one can also summarize the data in the frequency domain by computing what is called the **cross power spectral density** (or **cross-spectral density**, CSD). This allows for a compact representation of long time series measurements and is typically done with "resting-state" data, *i.e.*, a situation without any experimentally controlled stimuli in which brain activity consists of "spontaneous" oscillations. This is the scenario we will consider in the following.

The CSD can be seen as a relationship between two time series as a function of frequency. More specifically, one can think of the CSD as a frequency domain analysis of the covariance between two signals. The CSD $S_{y_i y_j}(\omega)$ between two signals $y_i(t)$ and $y_j(t)$ is given by:

$$S_{y_i y_j}(\omega) = \mathbb{E}[\mathcal{F}\{y_i\}\overline{\mathcal{F}\{y_j\}}] \tag{18}$$

where $\mathcal{F}\{y_i\}$ is the Fourier transform of signal $y_i(t)$ and $\overline{\mathcal{F}\{y_i\}}$ the complex conjugate of $\mathcal{F}\{y_i\}$.

For the special case $i = j$, $S_{y_i y_j}(\omega)$ is called the **power spectral density** (PSD).

Since electrophysiological measurements are performed using several sensors, we obtain as many time-series as we have channels[6]. If one computes the CSD for each possible pair of signals $y_i(t)$ and $y_j(t)$, one obtains a symmetric matrix for each considered frequency $\omega$. This means that the full CSD, for all considered frequencies, is a three-dimensional tensor (Figure 8).

Consequently, if we aim to model data in the frequency domain, we need to add yet another step to the generative model that maps time-series data (Equation 17) to CSD, according to Equation 18. For a linear system, this is done most efficiently by expressing the mapping between the so-called neuronal fluctuations (or endogenous oscillations; recall that we are in a resting-state scenario) and the measured signal in terms of the kernel of this system [54]. As we will see in the next section, in this case the forward mapping simply corresponds to a convolution of the neuronal fluctuations with the derived kernel.

---

[6] While this is true for the raw (*i.e.*, unprocessed) data, in practice, principal component analysis (PCA) is often used as a preprocessing step to reduce the dimensionality of the data.

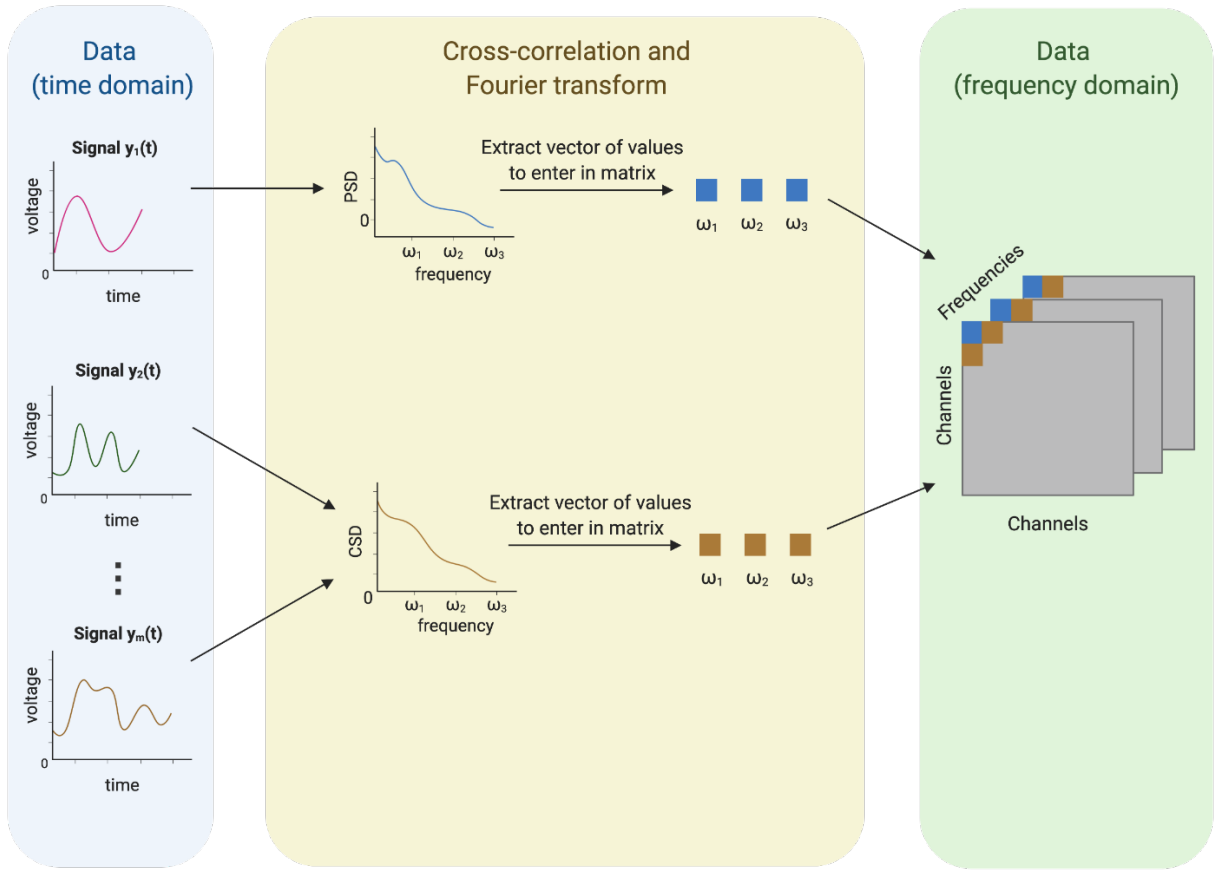

*Figure 8: Expressing time-domain data in the frequency domain. CSD = cross spectral density; PSD = power spectral density. Figure created with Biorender.com*

## 9 Modeling cross-spectral densities

In this section, we will specify a generative model of predicted CSD for resting-state data, based on a mapping between neuronal fluctuations or innovations $u_k(t) \in \mathbb{R}$ (which can be thought of as baseline oscillations of neuronal signal driving the neural populations[7]) and observable signals $y_i(t) \in \mathbb{R}$. Under linear assumptions, this mapping is defined by convolving the neuronal innovations with their corresponding kernel $\kappa_i^k(\tau, \theta)$. A kernel $\kappa_i^k(\tau, \theta)$ is specified as follows [54]:

$$\kappa_i^k(\tau, \theta) = \frac{\partial y_i(t)}{\partial u_k(t-\tau)} \tag{19}$$

---

[7] Refer to Section 11 for a detailed account of how these fluctuations are modeled in SPM12.

Here, $\theta$ are the parameters we wish to estimate, $k$ indicates the $k$-th innovation and $i$ the $i$-th channel [54]. Note that $u_k(t - \tau)$ is defined with a time lag $\tau$. Indeed, there is a delay between the generation of inputs $u_k(t - \tau)$ and that of the signal $y_i(t)$.

Finally, note that, for ease of writing, the Fourier transform will now be indicated by a capital letter (*e.g.*, $K_i^k = \mathcal{F}\{\kappa_i^k\}$). For further simplicity, we follow the notation from Friston *et al.* [54] and omit the dependencies on the frequency $\omega$ and parameters $\theta$ in the following derivations. The predicted CSD are denoted as $g_{ij}(\omega, \theta)$. According to Equation 18, the predicted CSD $g_{ij}(\omega, \theta)$ is formulated as follows:

$$\begin{aligned}
g_{ij}(\omega, \theta) &= \mathbb{E}\left[\mathcal{F}\{y_i\} \cdot \overline{\mathcal{F}\{y_j\}}\right] \\
&= \mathbb{E}\left[\mathcal{F}\left\{\sum_k \kappa_i^k * u_k\right\} \cdot \overline{\mathcal{F}\left\{\sum_l \kappa_j^l * u_l\right\}}\right] \\
&= \mathbb{E}\left[\sum_k \sum_l K_i^k \cdot U_k \cdot \overline{K_j^l \cdot U_l}\right] \\
&= \mathbb{E}\left[\sum_k \sum_l K_i^k \cdot \overline{K_j^l} \cdot U_k \cdot \overline{U_l}\right] \\
&= \sum_k \sum_l K_i^k \cdot \overline{K_j^l} \cdot \mathbb{E}[U_k \cdot \overline{U_l}]
\end{aligned} \tag{20}$$

It is therefore assumed that the CSD can be represented as a sum of convolved neuronal innovations. In the last step, we use the fact that the expectation operator is linear and that $U_k$ and $U_l$ are our random variables of interest (in the frequency domain). In addition, as will be seen later, since the equations of motion are linear in $u_k(t)$, the kernel does not depend on the neuronal innovations, therefore allowing us to pull the first two terms out of the expectation.

The real and imaginary parts of $U_k$ are assumed to be identically and independently distributed (i.i.d.), as follows [54]:

$$p(Re(U_k)) = \mathcal{N}(0, \gamma_k) \tag{21}$$

$$p(Im(U_k)) = \mathcal{N}(0, \gamma_k) \tag{22}$$

Note that if two random variables $X$ and $Y$ are independent, then their covariance is zero. This means that: $cov(X, Y) = \mathbb{E}[(X - \mathbb{E}(X)(Y - \mathbb{E}[Y])] = 0$. Applying this standard definition of the covariance, we obtain the following relation:

$$\begin{aligned}
cov(Re(U_k), Re(U_l)) &= \mathbb{E}[Re(U_k) - \mathbb{E}[Re(U_k)]] \cdot \mathbb{E}[Re(U_l) - \mathbb{E}[Re(U_l)]] \\
&= \mathbb{E}[Re(U_k) - 0] \cdot \mathbb{E}[Re(U_l) - 0] \\
&= \mathbb{E}[Re(U_k)] \cdot \mathbb{E}[Re(U_l)] \\
&= 0
\end{aligned} \tag{23}$$

The same applies to the imaginary component of $U_k$:

$$cov(Im(U_k), Im(U_l)) = \mathbb{E}[Im(U_k)] \cdot \mathbb{E}[Im(U_l)]$$

$$= 0 \qquad (24)$$

In addition, note that, using the linearity of the expectation operator, the expression for $\mathbb{E}[U_k \cdot \overline{U_l}]$ can be rewritten as:

$$\begin{aligned} \mathbb{E}[U_k \cdot \overline{U_l}] &= \mathbb{E}\big[(Re(U_k) + j \cdot Im(U_k))(Re(U_l) - j \cdot Im(U_l))\big] \\ &= \mathbb{E}[Re(U_k) \cdot Re(U_l)] - j \cdot \mathbb{E}[Re(U_k) \cdot Im(U_l)] \end{aligned}$$

$$+ j \cdot \mathbb{E}[Im(U_k) \cdot Re(U_l)] + \mathbb{E}[Im(U_k) \cdot Im(U_l)] \qquad (25)$$

where $j$ denotes the imaginary number: $j = \sqrt{-1}$.

Thus, if $k \neq l$ :

$$\mathbb{E}[U_k \cdot \overline{U_l}] = 0 \qquad (26)$$

since the aforementioned i.i.d. assumption also implies that the cross-terms from Equation 25 are zero.

If $k = l$ :

$$\begin{aligned} \mathbb{E}[U_k \cdot \overline{U_k}] &= \gamma_k + 0 + 0 + \gamma_k \\ &= 2\gamma_k \end{aligned}$$

$$= \lambda_k \qquad (27)$$

where we define a new variable $\lambda_k$ to be the spectral density of the neuronal innovations.

Therefore, making use of Equations 26 and 27, Equation 20 simplifies to:

$$\begin{aligned} g_{ij}(\omega, \theta) &= \sum_k \sum_l K_i^k \cdot \overline{K_j^l} \cdot \mathbb{E}[U_k \cdot \overline{U_l}] \\ &= \sum_k K_i^k \cdot \overline{K_j^k} \cdot \mathbb{E}[U_k \cdot \overline{U_k}] \\ &= \sum_k K_i^k \cdot \overline{K_j^k} \cdot \lambda_k \\ &= \sum_k g_{ij}^k(\omega, \theta) \end{aligned} \qquad (28)$$

One hence sees that the predicted cross-spectrum is a linear mixture of the cross-spectra induced by each innovation [54]. Importantly, rather than making assumptions about the neuronal innovations in the time domain, one can instead directly parameterize the spectral density of the neuronal innovations, $\lambda_k$. For this, the following form has been suggested based on previous theoretical and experimental work [73]–[76]:

$$\lambda_k(\omega) = \alpha + \frac{\beta^{(1)}}{\omega^{\beta^{(2)}}} \tag{29}$$

Here, $\alpha$ represents white noise and the second term colored noise, with $\beta^{(2)}$ indicating the "color" of the noise [8] and $\beta^{(1)}$ its magnitude. In the most recent version of SPM, this is implemented somewhat differently. We will review precisely how SPM models the spectral density of the neuronal innovations in Section 11 of this paper.

**Specifying the kernel**

To be able to generate predictions, we now need to specify the kernels $\kappa_i^k(\tau, \theta)$. These can be computed analytically under further simplifying assumptions. We first define the forward mapping which links endogenous innovations $u$ to hidden states $x$ and finally to the observed signal $y_i$ (for channel $i$). This mapping contains both the equations of motion $\dot{x} = f(x(t), \theta, u(t))$ and the observation model $g(x(t), \theta)$. For notational simplicity, we omit in the following derivations the dependency of both these functions on $\theta$, $u$ and $x$, thus expressing them as $f(t)$ and $g(t)$, respectively. Furthermore, we define $J = \partial f / \partial x$ as the Jacobian of the neuronal system. Using the chain rule [54]:

$$\kappa_i^k(\tau, \theta) = \frac{\partial y_i(t)}{\partial u_k(t-\tau)}$$

$$= \frac{\partial y_i(t)}{\partial g(t)} \cdot \frac{\partial g(t)}{\partial x(t)} \cdot \frac{\partial x(t)}{\partial x(t-\tau)} \cdot \frac{\partial x(t-\tau)}{\partial \dot{x}(t-\tau)} \cdot \frac{\partial \dot{x}(t-\tau)}{\partial u_k(t-\tau)} \tag{30}$$

By assuming a linear differential equation with Jacobian $J = \partial \dot{x}/\partial x$, we have: $\dot{x}(t) = J \cdot x(t)$

Therefore: $x(t) = \exp(Jt) \cdot C$, where $C$ represents the initial conditions of this system.

One can hence re-express this equation as: $x(t) = \exp(J\tau) \cdot x(t-\tau)$

Thus, by inserting this last expression into Equation 30:

$$\kappa_i^k(\tau, \theta) = 1 \cdot \frac{\partial g(t)}{\partial x(t)} \cdot \exp(J\tau) \cdot \left(\frac{\partial \dot{x}(t-\tau)}{\partial x(t-\tau)}\right)^{-1} \cdot \frac{\partial \dot{x}(t-\tau)}{\partial u_k(t-\tau)}$$

---

[8] As can be seen in Equation 29, in colored noise, the power spectral density is, for positive $\beta^{(2)}$, inversely proportional to the frequency $\omega$ of the signal [82].

$$= \frac{\partial g(t)}{\partial x(t)} \cdot \exp(J\tau) \cdot \left(\frac{\partial \dot{x}(t)}{\partial x(t)}\right)^{-1} \cdot \frac{\partial \dot{x}(t)}{\partial u_k(t)}$$

$$= \frac{\partial g(t)}{\partial x(t)} \cdot \exp(J\tau) \cdot J^{-1} \cdot \frac{\partial f(t)}{\partial u_k(t)} \tag{31}$$

Note that the omission of the delay $\tau$ from the first to the second line is justified by the fact that $\partial \dot{x}(t)/\partial x(t)$ is linearly approximated and that the equations of motion $\dot{x}(t)$ are linear with respect to the inputs. Thus, taking the derivative at one or a later time point is assumed to yield the same result. In addition, because of the second linearity, note that the kernel expression does not depend on $u_k(t)$.

**Adding noise**

To complete the specification of the forward mapping to cross-spectral data, it is presumed that the data is a mixture of the output of the forward mapping (*i.e.,* predicted cross-spectra $g_{ij}(\omega, \theta)$ and channel noise $\lambda^c(\omega)$ ) and Gaussian error $\epsilon_{ij}(\omega)$ [54]:

$$g_{ij}(\omega) = g_{ij}(\omega, \theta) + \lambda^c(\omega) + \epsilon_{ij}(\omega)$$

$$= \sum_k K_i^k \cdot \overline{K_j^k} \cdot \lambda_k + \lambda^c(\omega) + \epsilon_{ij}(\omega) \tag{32}$$

Where it is assumed that:

$$\lambda^c(\omega) = \eta + \frac{\nu^{(1)}}{\omega^{\nu^{(2)}}} \tag{33}$$

$$\text{Re}(\epsilon_{ij}) \sim \mathcal{N}(0, \Pi(\omega)_\epsilon^{-1}) \tag{34}$$

$$\text{Im}(\epsilon_{ij}) \sim \mathcal{N}(0, \Pi(\omega)_\epsilon^{-1}) \tag{35}$$

The spectral density of the channel noise $\lambda^c(\omega)$ is parameterized using a white ($\eta$) and a colored component. Once again, this is implemented somewhat differently in the most recent version of SPM and the exact procedure will be reviewed in the Section 11.

As a final remark, note the conceptual differences between the channel noise and Gaussian error terms: channel noise represents structured noise which is integrated in the forward mapping, and contains parameters which are estimated during model inversion. On the other hand, Gaussian error should be interpreted as measurement noise. This term furnishes the form of the likelihood function (see next section) and is defined by hyperparameters.

## 10 Model inversion

In the previous sections, we discussed how the forward model of cbDCM for cross-spectral densities is constructed, *i.e.*, how data could be generated from a set of parameters. Typically, we however face the inverse challenge: given some measured data $y$, how can one obtain the most probable parameter distribution which could have generated the given data? In the context of the generative model, solving this inverse problem is called model inversion or inference. In DCM, this is achieved within a Bayesian framework. In this setting, for a given model $m$, parameter estimation is equivalent to computing the moments of the posterior distribution of the parameters $\theta$ given the observed data $y$ [25], according to Bayes' rule:

$$p(\theta|y,m) = \frac{p(y|\theta,m)p(\theta|m)}{p(y|m)} \quad \text{where:} \quad p(y|m) = \int p(y|\theta,m)p(\theta|m)\,d\theta \tag{36}$$

According to Equation 36, computing the posterior $p(\theta|y,m)$ requires the likelihood $p(y|\theta,m)$, the prior $p(\theta|m)$ and the normalization constant $p(y|m)$. The forward mapping from hidden states to observable signals described in the previous section defines the likelihood function. In addition, informed priors can be specified over the parameters (for examples, see [49]). However, because the normalization constant (or model evidence) $p(y|m)$ is an integral which can be very difficult to compute, calculating the posterior distribution of the parameters can be seldom done analytically. Instead, model inversion typically relies on approximate inference. One such approximate procedure is Variational Bayes (VB). VB turns the aforementioned integration problem into an optimization problem, which involves maximizing a quantity termed the negative free energy $F(q,y)$[9]:

$$F(q,y) = \underbrace{\langle \log p(y|\theta,m) \rangle_q}_{\text{Expected log-likelihood}} - \underbrace{D_{KL}[q(\theta)\,||\,p(\theta|m)]}_{\text{Kullback–Leibler divergence}} \tag{37}$$

$F(q,y)$ represents a lower-bound approximation to the logarithm of the model evidence under an approximate posterior $q(\theta)$, which represents our best guess of the form the posterior may have. In DCM, the approximate posterior $q(\theta)$ is chosen to be a Gaussian [52]. Under this approximation, model inversion of DCMs rests on updating the sufficient statistics of $q(\theta)$ (*i.e.,* mean and covariance) in order to maximize the negative free energy. After convergence, $F(q,y)$ represents our best approximation to the log model evidence and, simultaneously, $q(\theta)$ will be the best approximation to the true posterior $p(\theta|y,m)$ (where "best" is to be understood in reference to the chosen form of $q(\theta)$).

Equation 37 shows that the negative free energy comprises two distinct terms. Specifically, the expected log-likelihood can be thought of as an accuracy term whereas the Kullback-Leibler (KL) divergence between approximate posterior and prior represents a complexity term (also known as Bayesian surprise) [77]. The complexity term effectively serves as a regularization term which acts against overfitting. More details on the conceptual basis of these methods, as well as the exact approach used for DCMs can be found in separate publications [35], [36].

---

[9] In the statistics and machine learning literature, $F(q,y)$ is often called the evidence lower bound (ELBO) [83].

## 11  Implementation in SPM

SPM is a free and open source MATLAB software package developed at the Functional Imaging Laboratory (FIL), Wellcome Trust Centre for Neuroimaging, London [33]. It contains a set of tools to analyze neuroimaging data (*e.g.*, fMRI, EEG, MEG) and can also be used to specify and estimate DCMs. Since the original papers introduced the theoretical underpinnings of the models, various additions and changes have been made in the SPM code. We here provide an overview of some of the relevant aspects of the current implementation of (conductance-based) DCM in SPM. Notably, this section does not attempt to contain an exhaustive list of all changes. Instead, we explore only those aspects that we deem most relevant for the user.

**Choosing your model**

If you use the SPM graphical user interface (GUI) and try to define a DCM, you will see a window similar to the one in Figure 9. You will be confronted with many options for both the observation and the neuronal models.

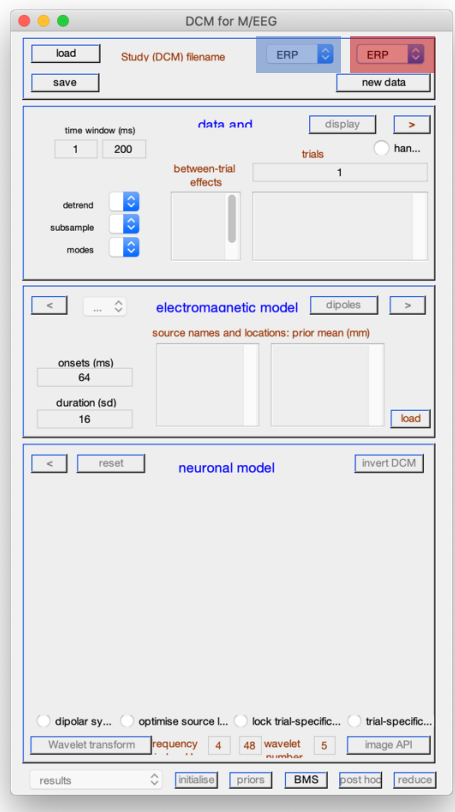

*Figure 9: SPM GUI. The dropdown menu within the blue box will allow you to choose an observation model, whereas the menu in the red box will permit you to select a neuronal model.*

To facilitate usage and provide guidance, we have listed the available options in SPM12 in Table 2 and Table 3 and explain their meaning with regard to the framework introduced in Section 2 of this paper.

*Table 2: SPM options for the **neuronal** model. Verification of which functions are called by which options can be done by consulting the spm_dcm_x_neural.m function.*

| Option | Neuronal model selected |
| --- | --- |
| ERP (evoked-response potential) | 3-population convolution-based NMM. |
| SEP (sensory evoked potential) | Variant of the ERP model, with different fixed parameter values (see spm_fx_sep.m). As with the ERP model, this is a 3-population convolution-based NMM. |
| LFP (local field potential) | 3-population convolution-based NMM. This model is conceptually the same as the ERP model, except for the inhibitory within-source connections: the inhibitory population has recurrent self-connections. These were included to allow for modeling of high-frequency oscillations in the beta band [78]. |
| CMC (canonical microcircuit) | 4-population convolution-based NMM |
| NMM (neural mass model) | 3-population conductance-based NMM. |
| MFM (mean field model) | 3-population conductance-based MFM. |
| CMM (canonical microcircuit model) | 4-population conductance-based NMM. |
| NMDA | 3-population conductance-based NMM (defining an MFM model is also possible, but not via the GUI). This model includes the NMDA receptor. |
| CMM_NMDA | 4-population conductance-based NMM. This model includes the NMDA receptor. |
| NFM (neural field model) | 3-population convolution-based NFM. Although possible via the GUI, combining this option along with an ERP or CSD observation model does not make sense and will consequently return an error (see spm_dcm_x_neural.m). |

*Table 3: SPM options for the **observation** model. Verification of which functions are called by which options can be done by consulting the spm_api_erp.m function.*

| Option | Neuronal model selected |
| --- | --- |
| ERP (evoked-response potential) | Observation model for ERPs. Returns predictions in the time domain. Can be used with all of the above neuronal models, except the NFM. |
| *CSD (cross-spectral density)* | Observation model for resting-state data. Returns predictions in the frequency domain. Can be used with all of the above neuronal models, except the NFM. |
| *TFM (time-frequency model)* | DCM of induced cross-spectra. SPM will automatically choose a CMC neuronal model for you (see spm_dcm_tfm.m and spm_fx_cmc_tfm.m)). |
| *IND (induced responses)* | DCM of induced responses. SPM will automatically choose the associated neuronal model for you (see spm_dcm_ind.m and spm_fx_ind.m). This model is not covered in the present publication. For details, refer to [46]. |
| *PHA (phase coupling)* | DCM for phase coupling. SPM will automatically choose the associated neuronal model for you (see spm_dcm_phase.m and spm_fx_phase.m). This model is not covered in the present publication. For details, refer to [44]. |
| *NFM (neural field model)* | Can only be used with the NFM neuronal model. |

**Modeling intrinsic connectivity**

The intrinsic connectivity rules for the conductance-based models implemented in the current version of SPM differ from what was initially described by David *et al.* [25]. Figure 10 and Figure 11 illustrate the within-source connectivity patterns currently implemented in SPM, with reference to the relevant functions.

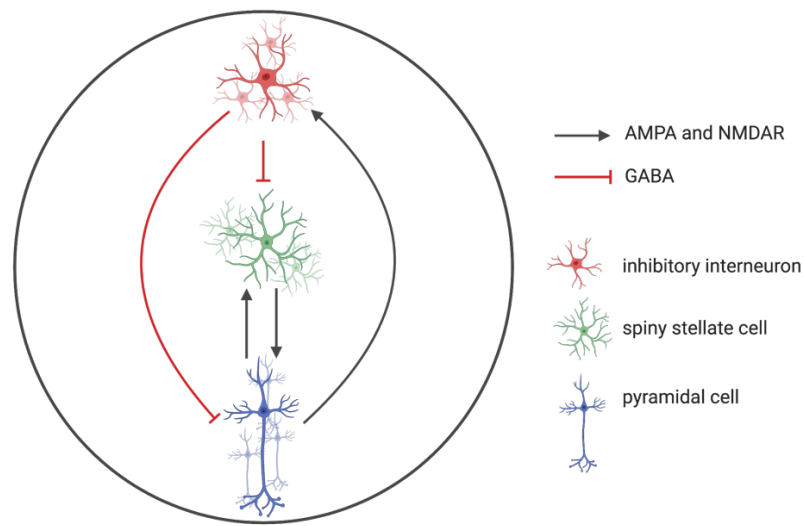

*Figure 10: Within-source connectivity for the 3-population source model, as implemented in the spm_fx_mfm.m and spm_fx_nmda.m functions. Figure created with Biorender.com*

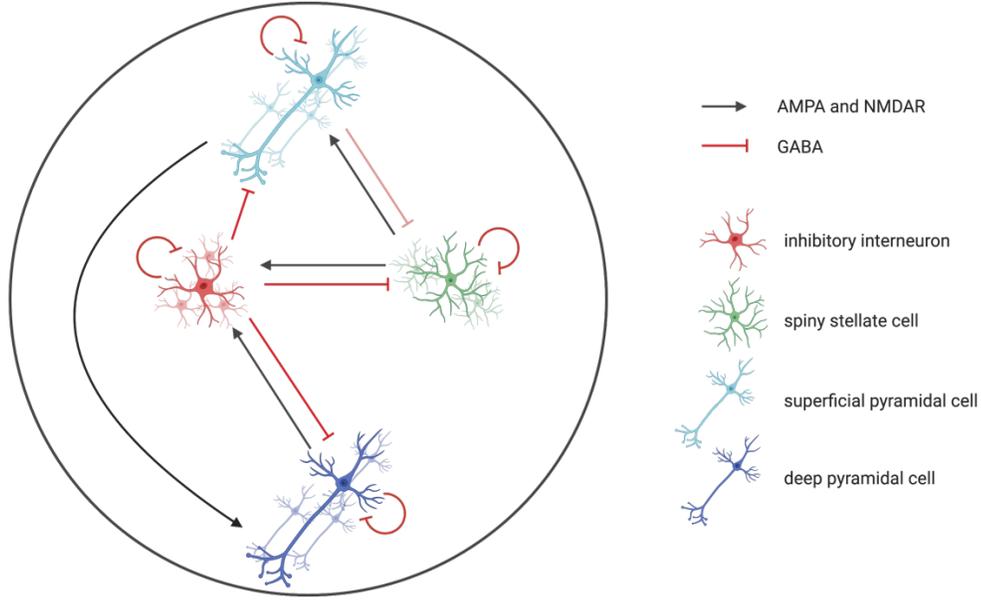

*Figure 11: Within-source connectivity for the 4-population source model, as implemented in the spm_fx_cmm.m and the spm_fx_cmm_NMDA.m functions. In the latter function, the inhibitory connection from the superficial pyramidal cell population to the spiny stellate cells is not represented (this is indicated by a less strong line in the figure). Figure created with Biorender.com*

**Modeling the neuronal innovations**

Within SPM, the spectral density of the neuronal innovations $\lambda_k$ is modeled somewhat differently from what was discussed previously (Equation 29). In SPM, 2 parameter classes are used to this effect: **a** and **d** (see spm_ssr_priors.m and spm_csd_mtf_gu.m). **a** models $1/\omega$ (colored) noise, whereas **d** contains the coefficients for a discrete cosine transform (DCT). This DCT is, to our knowledge, not mentioned explicitly in the literature as a model of endogenous fluctuations in DCM for cross-spectral data; however, it is implemented in the more recent versions of SPM.

In SPM, **a** is defined as a matrix:

$$\mathbf{a} = \begin{bmatrix} a_{11} & a_{12} & \cdots & a_{1n} \\ a_{21} & a_{22} & \cdots & a_{2n} \end{bmatrix} \tag{38}$$

where $n$ represents the number of sources in the DCM. As for the first dimension of **a**, for each source $i$, the colored noise ($1/\omega$) component $\lambda_a^{(i)}(\omega)$ is described as:

$$\lambda_a^{(i)}(\omega) = \exp(a_{1i}) \cdot \omega^{-\exp(a_{2i})} \tag{39}$$

$\lambda_a^{(i)}(\omega)$ is then a scalar value. If one does this operation for all sources, one obtains the $1 \times n$ vector $\lambda_a(\omega)$ for each frequency $\omega$:

$$\lambda_a(\omega) = \begin{bmatrix} \lambda_a^{(1)}(\omega) \\ \lambda_a^{(2)}(\omega) \\ \vdots \\ \lambda_a^{(n)}(\omega) \end{bmatrix}^T \qquad (40)$$

**d** is also a matrix, of the following form:

$$\mathbf{d} = \begin{bmatrix} d_{11} & d_{12} & \cdots & d_{1n} \\ d_{21} & \ddots & \ddots & d_{2n} \\ \vdots & \ddots & \ddots & \vdots \\ d_{41} & d_{42} & \cdots & d_{4n} \end{bmatrix} \qquad (41)$$

Note that the first dimension indicates the number of cosine functions used to model $\lambda_k$ and that it can be arbitrarily defined. In the current version of SPM, this number has been set to 4, while previous versions worked with 8 cosine functions[10]. Once again, the parameters are defined in a source-specific manner. The DCT component $\lambda_d(\omega)$ of the spectral density of the neuronal innovations is modeled for all sources using matrix multiplication.

$$\lambda_d(\omega) = \exp(C(\omega) \times d) \qquad (42)$$

Here, $C(\omega)$ is a row of the DCT matrix and $\exp()$ indicates the elementwise exponential. Therefore, $\lambda_d(\omega)$ also constitutes a $1 \times n$ vector for each frequency $\omega$.

The $1/\omega$ and DCT components are then combined using an elementwise product:

$$\lambda(\omega) = \lambda_a(\omega) \circ \lambda_d(\omega) \qquad (43)$$

where $\lambda(\omega)$ corresponds to the vector of neuronal innovations indexed by $k$ in Equation 32.

**Modeling noise**

We now address the channel noise terms introduced in Equations 32 and 33. In SPM, the colored channel noise added to the predicted CSD is subdivided into what is called non-specific channel noise (which describes the contribution of common noise sources, *e.g.*, in a common reference channel) and specific channel noise [79]. The white noise component $\eta$ (Equation 33) is set to zero; yet, white noise can, in principle, still be represented, as we will see below. Mathematically, the model is expressed as follows:

---

[10] Note that this DCT is modeling frequency-domain (and not time-series) data. Thus, DCT frequencies cannot be equated to the frequencies of the time-domain data. Furthermore, note that the DCT is a data-driven process. Thus, the range of frequencies picked up by the DCT depend on the characteristics of the data.

$$g_{ij}(\omega) = \sum_k K_i^k \cdot \overline{K_j^k} \cdot \lambda_k + g_{ij}^{noise}(\omega) + \epsilon_{ij}(\omega) \tag{44}$$

Here, we have substituted $\lambda^c(\omega)$ with $g_{ij}^{noise}(\omega)$. In SPM, two parameter classes model channel noise: **b** and **c**. Each of these parameter classes is a $2 \times 1$ vector:

$$\mathbf{b} = \begin{bmatrix} b_1 \\ b_2 \end{bmatrix} \quad \mathbf{c} = \begin{bmatrix} c_1 \\ c_2 \end{bmatrix} \tag{45}$$

where **b** contains the parameters that model non-specific channel noise, whereas **c** models specific channel noise. $g_{ij}^{noise}(\omega)$ is then expressed as:

$$g_{ij}^{noise}(\omega) = \begin{cases} \exp(b_1 - 2) \cdot \omega^{-\exp(b_2)} + \exp(c_1 - 2) \cdot \omega^{-\exp(c_2)} & \text{if: } i = j \\ \exp(b_1 - 2) \cdot \omega^{-\exp(b_2)} & \text{otherwise} \end{cases} \tag{46}$$

Note that **b** and **c** are not single-channel-specific and that, in the extreme case where $b_2$ or $c_2$ take very small negative numbers, both $-\exp(b_2)$ and $-\exp(c_2)$ will be close to zero and $g_{ij}^{noise}(\omega)$ will essentially model white noise.

## 12 Simulations

So far, we have reviewed the mathematical equations for cbDCM for cross-spectral densities derived from electrophysiological data. In addition, we discussed some aspects of the model implementation in SPM that we deemed useful for the reader. In this section, we conclude this second, more practical part of the paper by making use of simulations to provide an intuitive and qualitative understanding of the role certain parameters play in the generation of cross-spectra.

To do this, we define two simple models (Figure 12) of cross-spectral LFP data. Both models are cbDCMs with four neural populations per source and the NMDA receptor. The first model contains only one source, representing the medial prefrontal cortex (mPFC), to illustrate the effect of within-source parameters. The second model includes two sources, representing the mPFC and the posterior cingulate cortex (PCC), to illustrate the effect of between-source parameters. Specifically, in the second model, one single extrinsic connection is defined, from the PCC to the mPFC.

To examine the effect of changing specific parameter values on the CSD, we start by taking the prior mean, as defined in SPM12, as parameter value to generate data. We then change the value of single parameters up to two prior standard deviations away from the prior mean. In addition, in one set of simulations, we change the current parameterization in SPM12 to illustrate the effect of the magnesium nonlinearity. While the magnesium block parameters are fixed by default in the current SPM implementation, they have been parameters of interest in several previous studies. For example, Moran *et al.* [50] parameterized the magnesium switch as in Equation 6. We also define $\alpha_{NMDA}$ as a parameter of interest and set it to values similar to those previously tested in this publication.

In the following, we plot the synthetic data over the full frequency band (1-48 Hz). Please note that the choice of parameters tested is somewhat arbitrary and that we deliberately defined simple models. Our aim is to provide the reader with an intuition of the dynamics of the models and not to make precise statements about what cbDCM can or cannot do. The code for these simulations is freely available and can be accessed here: https://gitlab.ethz.ch/tnu/code/pereiraetal_conductance_based_dcm. This is meant to offer an easy, hands-on experience with cbDCMs and we encourage the reader to experiment with the code.

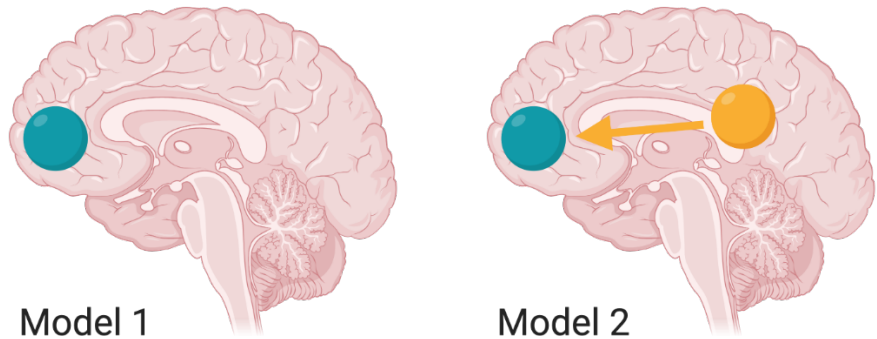

*Figure 12: Models used in the simulations. On the left, model 1 includes one single source, representing the mPFC. On the right, model 2 contains 2 sources: the mPFC anteriorly, and the PCC posteriorly. These sources are coupled via one single extrinsic forward connection, from PCC to mPFC. Figure created with Biorender.com*

**Single-source model**

We start by analyzing the parameters of the neuronal model. Figure 13 shows the results of the simulations when one changes the values of four different parameters (please note that the y-axes of the plots are scaled differently). In the context of this first single-source LFP model, we assume only one electrode is present, and therefore have only one channel (Figure 8), and only one plot per parameter tested.

Let us start with $\kappa_{AMPA}$, the AMPA receptor rate constant (Equation 8, subplot A in Figure 13). Visually, one sees that changes in this parameter's value lead to more apparent alterations in the peak around 20 Hz, with decreases in the value of $\kappa_{AMPA}$ leading to an increase in the magnitude of the PSD. Subplot B is similar to the previous plot, with the difference that the parameter changed is the within-source coupling parameter $\gamma_k^{(j,i)}$ (Equation 15), weighting the connection from the spiny stellate cells ($i$) to the superficial pyramidal cells ($j$). This particular connection was chosen because the superficial pyramidal cells constitute the neural population that is assumed (within SPM12 for 4-population conductance-based models) to most significantly contribute to the measured signal (for a detailed discussion on the contribution of the different populations, refer to [80]). Increases in the strength of this connection would intuitively lead to an increase in the magnitude of the spectra, which is what is observed in Figure 13. This effect is much more pronounced for frequencies below 20 Hz.

Parameter $\Sigma^{(j)}$ (Equation 15) is assumed to take on the same value for all neuronal populations in the model. When testing this parameter, we obtained an effect qualitatively similar to that observed for $\kappa_{AMPA}$ (subplot C). Finally, we turn to $\alpha_{NMDA}$, which exerts a profound influence on the PSD, and determines which frequency band is associated with the maximal magnitude. In the current implementation of SPM12, $\alpha_{NMDA}$ is fixed to 0.06 (yet has been used as a free parameter in various previous publications [14], [50]). In our simulations (subplot D), values below 0.06 are associated with maximal PSD in the lower frequencies (< 5 Hz). Conversely, as the value of $\alpha_{NMDA}$ is increased, a significant rise in the value of the spectra in the 15 to 25 Hz interval occurs, with the beta band (13-29 Hz) eventually becoming the band associated with the maximal PSD.

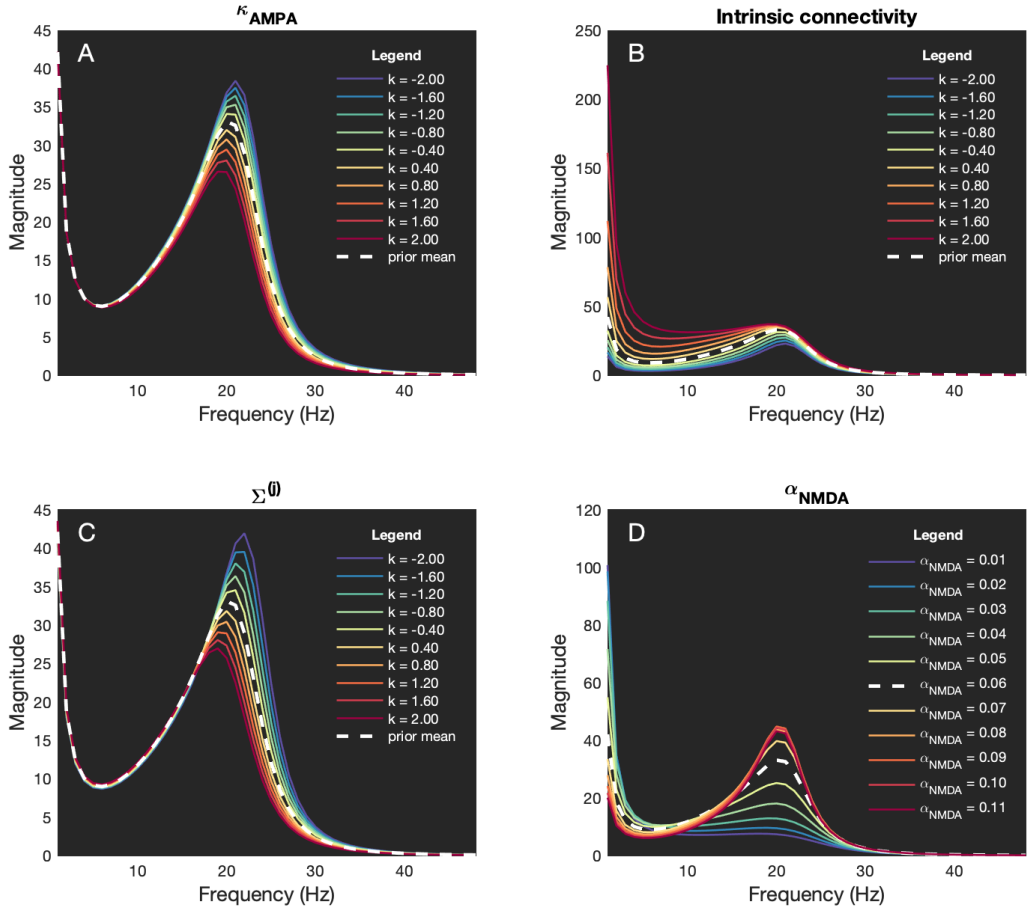

*Figure 13: Simulated data using model 1, as specified in Figure 12. The prior parameter values from SPM12 were used to perform the simulations of plots A, B and C. For these plots, the value of each parameter was altered up to two prior standard deviations away from the prior mean. $k = 2$ indicates that two standard deviations were added to the prior parameter value, whereas $k = -2$ signals that two standard deviations were removed. In addition, data were also simulated using the prior value for the parameter (which is equivalent to $k = 0$). Only the magnitude of the PSD is plotted. **(A)** Changes in the value of $\kappa_{AMPA}$ from source mPFC are more strongly reflected in the 15-30 Hz interval. **(B)** Increases in the weight of the connection from the spiny stellate cells to the superficial pyramidal cells gives rise to a predicted increase in the magnitude of the spectra, which is more evident below 20 Hz. **(C)** Changes in the value of parameter $\Sigma^{(j)}$ lead to alterations similar to those observed in subplot A. **(D)** Here, the magnesium switch was parameterized and $\alpha_{NMDA}$ was allowed to vary. The values tested fall within those evaluated in a previous publication [50], and the white dotted line corresponds to the default value in SPM12. $\alpha_{NMDA}$ plays an important role by determining which frequency band is associated with the maximal magnitude.*

Figure 14 displays the data generated by modifying the value of $a_2$, the exponent of the colored noise component of the neuronal innovations (Equation 39). Changes in $a_2$ lead to a qualitative effect again similar to that seen for $\kappa_{AMPA}$ and $\Sigma^{(j)}$, albeit with more pronounced alterations in the frequencies below 20 Hz.

Changes in the DCT component of the neuronal innovations also lead to interesting alterations. For this set of simulations, we change the value of a single DCT coefficient up to 5 prior standard deviations away from the prior mean, to make the qualitative effect more apparent. Figure 15 shows how changes in the value of the 4-th DCT coefficient for source mPFC leads to differences in the strength of an oscillation on top of the prior spectral density. In particular, with more extreme values of the DCT coefficient, one can observe more pronounced oscillations around the prior value ($k = 0$).

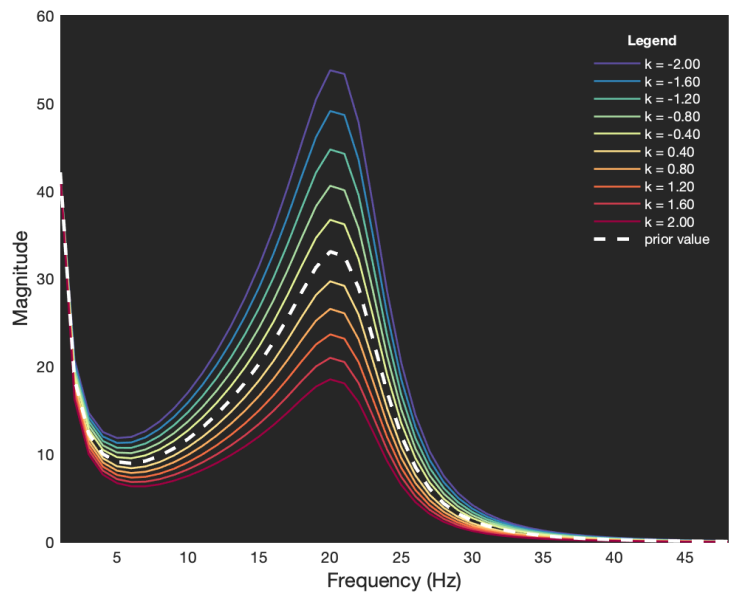

Figure 14: Simulated data, now changing the value of $a_2$, the exponent for the colored noise component of the neuronal innovations.

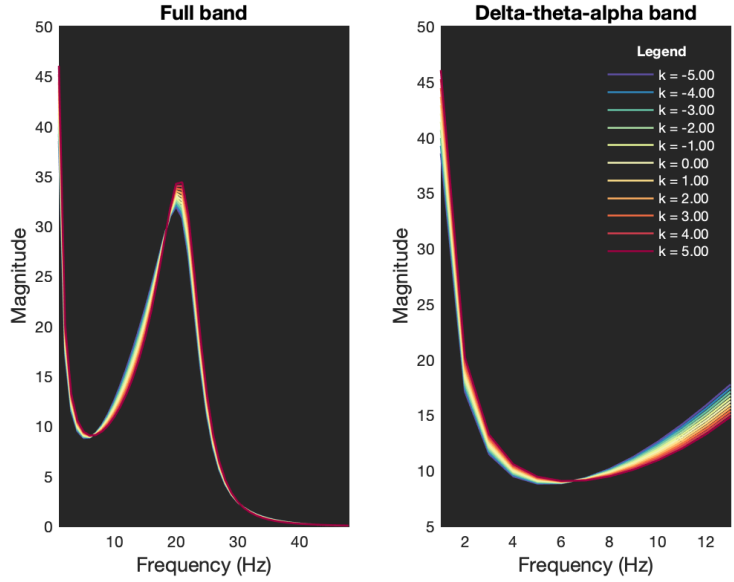

*Figure 15: Same as Figure 14, now changing the value of $d_{41}$, the 4-th DCT coefficient for source mPFC (DCT component of the neuronal innovations). The left plot shows the results over the full band, whereas the right plot "zooms in" on the lower frequencies (1-13 Hz), for a clearer visualization of the effect the parameter has on model output.*

Finally, we investigated the effect of the channel noise. Figure 16 shows the data simulated after changing the value of $c_2$, the exponent of the specific colored channel noise (Equation 46). It is evident that under our current settings, the changes induced in the power-spectra are subtle. By inspecting the full band, one does not see any changes. Even after "zooming in" on very high frequencies, the effect still appears very modest. This is likely due to the fact that, under the prior settings, the amplitude of the channel noise is given by $\exp(-2)$ (compare Equation 46).

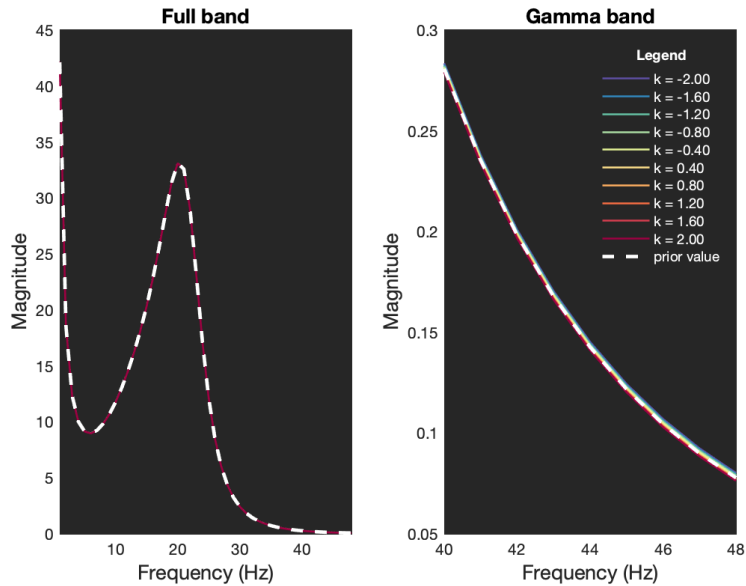

*Figure 16: Same as Figure 14, now changing the value of $c_2$, the exponent of the specific colored channel noise.*

**Two-source model**

The two-source model was defined specifically to study the effect of the extrinsic connectivity parameters on the power- and cross-spectral densities. As pointed out previously, this second model incorporates 2 sources: mPFC anteriorly, and PCC posteriorly. These are connected via one single forward projection from PCC to mPFC. Again, by modeling LFPs, we can assume one electrode per source, therefore having a total of two channels (Figure 8). In Figure 17, we simulate data after changing the value of the single AMPA-mediated extrinsic connectivity parameter. Only the upper half of the panels is shown, given the symmetry in the data. The magnitude of the complex spectra is plotted and the panel titles indicate which channels are being considered. As expected, the extrinsic connectivity parameter influences the scaling of the spectra, with higher values being associated with increased magnitude of the CSD. Note that the cross-spectra in panel CSD 2/2 are not affected by changes in this parameter. This is because we define no backward connection from the mPFC (channel 1) to the PCC (channel 2). Therefore, changes in the connection from the PCC to the mPFC will only elicit changes in the signal from the mPFC, and will not affect the PCC.

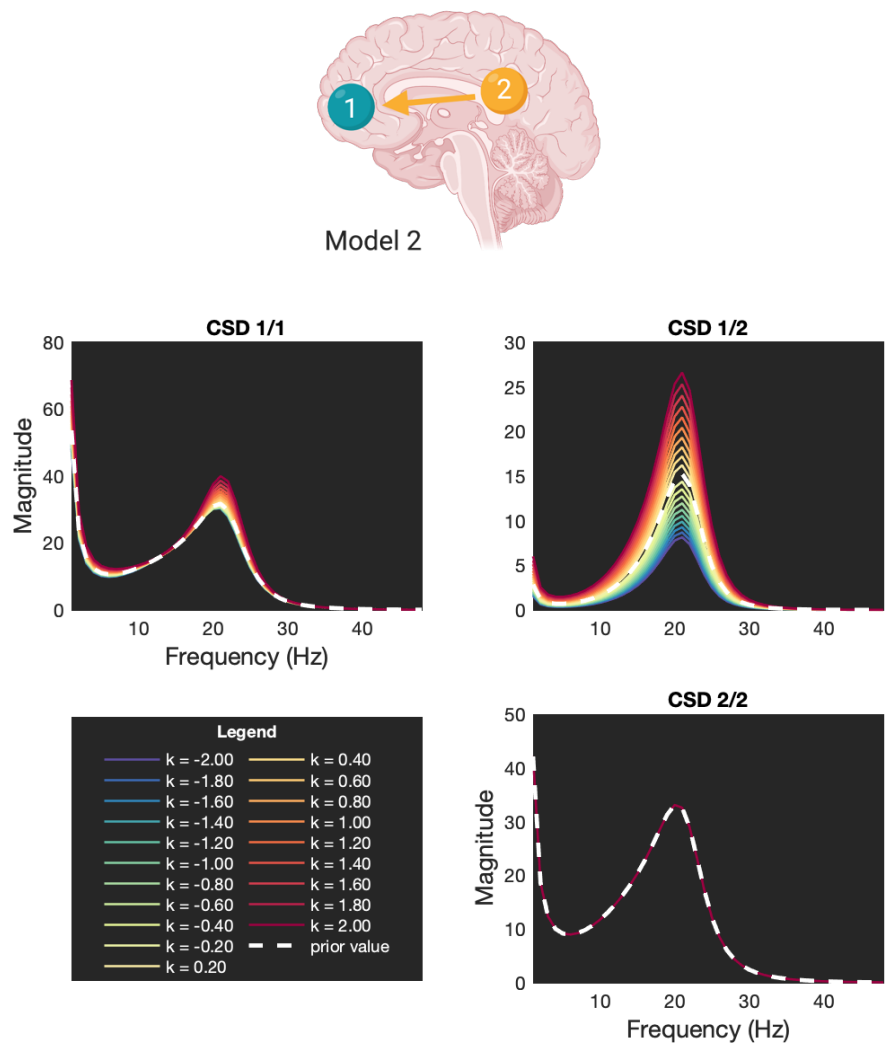

*Figure 17: Simulated data using model 2, as specified in Figure 12. The prior parameter values from SPM12 were once again used to perform the simulations. In this setting, the weight of the AMPA-mediated extrinsic connection from the PCC to the mPFC was altered, up to two prior standard deviations away from the prior mean. Only the upper half of the panels is shown, given the symmetry in the data. The panel titles indicate which channels are being considered. Only the magnitude of the cross-spectra is plotted. Figure created with Biorender.com*

To summarize, while all of the parameters are associated with changes in the model output, the qualitative and quantitative degree of change vary depending on the parameter considered. For instance, while $c_2$ exerted a very modest effect on the cross-spectra, other parameters, such as $a_2$, had a much more marked effect on the model output. In addition, certain parameters seem to preferably affect one frequency band over another. For instance, while the intrinsic connectivity parameter affected mostly the lower frequencies, other parameters, such as $\kappa_{\text{AMPA}}$, exerted a greater influence on the beta band. Finally, certain parameters showed an interesting qualitative effect, such as the DCT component of the neuronal innovations and, in particular, the magnesium nonlinearity parameter $\alpha_{\text{NMDA}}$.

## 13 Summary and conclusions

With the increasing use of more complex and sophisticated models in neuroscience, understanding the conceptual and mathematical principles behind these models has become more challenging. DCM for electrophysiological data, which was the focus of the present paper, builds on previous neurophysiological models and works with several layers of approximation. Keeping track of these steps is a difficult but crucial task if one is to understand the foundations of this class of generative models and be able to judge their strengths but also limitations.

In this paper, we have discussed the different DCM variants for electrophysiological data and reviewed cbDCM in particular. We have described the neuronal model in detail, explaining how single-neuron dynamics can be incorporated in a microcircuit model. We then turned to the observation model and derived it for resting-state electrophysiological data (CSDs).

In a second, more practical part of this paper, we reviewed several aspects related to the current implementation of these models in SPM. In particular, we discussed how intrinsic connectivity is represented for the 3- and 4-population neuronal models, and detailed how neuronal innovations and channel noise are modeled. Finally, we presented simulations from very simplistic models, in order to equip the reader with a qualitative understanding of how changes in specific parameters can alter the model's output.

We hope this tutorial paper will prove useful not only for readers starting to work with DCM for electrophysiological data, but also to scientists with more experience with these models.

## 14 Acknowledgements

This paper is based on the MSc thesis "DCM for anti-NMDAR encephalitis revisited – detailed model testing and simplification" by Inês Pereira at the University of Zurich and ETH Zurich [81]. Figures created with Biorender.com. KES is supported by the University of Zurich and the René and Susanne Braginsky Foundation.